\documentclass[
 reprint,
 superscriptaddress,
 amsmath,amssymb,
 aps,
 prb,
]{revtex4-2}

\usepackage[T1]{fontenc}
\usepackage{graphicx}
\usepackage{booktabs}
\usepackage{dcolumn}
\usepackage{bm}
\usepackage{xcolor}
\usepackage{hyperref}
\newcommand{\kdotp}{k·p }

\begin{document}

\title{High-fidelity k·p representations of first-principles electronic band structures through covariant renormalization}

\author{Kristian Berland}
\email{kristian.berland@nmbu.no}
\affiliation{Department of Mechanical Engineering and Technology Management, \\ Norwegian University of Life Sciences, NO-1432 Ås, Norway
}%

\date{\today}

\begin{abstract}
The \kdotp method can be applied directly to first-principles energies and
momentum matrix elements, but the resulting models converge slowly with the
number of bands and are inexact wherever the underlying Hamiltonian is
nonlocal. We show that both limitations can be largely removed by renormalizing
the eigenvalue spectra of the Hermitian momentum matrices. These spectra are
gauge invariant, and are identical along symmetry-related directions;
hence scaling the recurring magnitudes provides a modest set of parameters that
preserves degeneracies and crystal symmetry without needing to construct a
symmetry-adapted basis. 
We choose to renormalize the models against reference eigenvalues and band
velocities on rays out of a high-symmetry point.
For GaP, a 15-band model can reproduce the band structure in the near-gap reference regions within
a few meV. 
For zincblende and wurtzite AlN, similar accuracy requires 30- or 85-band models,
	which in the case of zincblende is traced to the strong warping in the [110] 
	direction.
The scheme readily generalizes to rocksalt PbTe, which includes spin-orbit coupling,
and has $L$-centered valence- and conduction-band extrema.
Further, we demonstrate, for GaP, how compact four-band Hamiltonians
can be obtained by downfolding within the same renormalization scheme. 
Moreover, by changing the fitting regime to encompass the entire Brillouin zone, the scheme can be used to generate full-zone models that reproduce the density of states.
A practical use of the method is illustrated with a 59-band \kdotp model for GaP, which can be evaluated on meshes far denser than the reference calculation, that in turn can be used both to resolve fine features in the density of states and to compute the hole conductivity at low temperature.
\end{abstract}
\maketitle

\section{Introduction}

The k$\cdot$p method is a standard method in solid-state physics.\cite{kaneBandStructureIndium1957, luttingerMotionElectronsHoles1955, willatzenMethodElectronicProperties2009}
It represents the electronic band structure in terms of the energies
$\varepsilon_n$ and generalized momentum matrix elements
$\bm{\pi}_{nm} = \langle \psi_{n\mathbf{k}_0} | \hat{\bm{\pi}} |
\psi_{m\mathbf{k}_0} \rangle$
between eigenstates at a reference wavevector $\mathbf{k}_0$,
typically a high-symmetry point in the Brillouin zone.
The generalized momentum operator $\hat{\bm{\pi}}$
is given by the wavevector derivative of the Hamiltonian,
\begin{equation}
\hat{\bm{\pi}} = \frac{m}{\hbar} \frac{\partial H(\mathbf{k})}{\partial \mathbf{k}}
 = m \hat{\mathbf{v}},
\label{eq:pi}
\end{equation}
which also equals $m$ times the velocity operator.
In the absence of spin-orbit coupling it reduces to the momentum
operator $\hat{\mathbf{p}}$, and $\bm{\pi}_{nm}$ to the momentum matrix
elements $\mathbf{p}_{nm}$.
In the basis of the eigenstates at $\mathbf{k}_0$, the Hamiltonian takes
the \kdotp form
\begin{equation}
H_{nm}(\mathbf{k}) = \varepsilon_n \delta_{nm}
 + \frac{\hbar}{m}\, \mathbf{q}\cdot\bm{\pi}_{nm}
 + \frac{\hbar^2 q^2}{2m}\, \delta_{nm},
\label{eq:bare-kp}
\end{equation}
with $\mathbf{q} = \mathbf{k}-\mathbf{k}_0$.
For a periodic local potential, the \kdotp method is formally exact, 
as eigenstates span the Hilbert space. 
In practice, \kdotp models are truncated, making the model less accurate with increasing
$\mathbf{q}$. 
The effects of these remote bands can be re-incorporated perturbatively using downfolding.
\cite{lowdinNoteQuantumMechanicalPerturbation1951, winklerQuasiDegeneratePerturbationTheory2003}

Traditionally, \kdotp models have been mostly constructed with a
handful of bands using a $\Gamma$-centered basis, such as for three valence orbitals and one conduction band ($| X \rangle$, $| Y \rangle$,
$| Z \rangle$, $|S \rangle$) for zincblende semiconductors. 
\cite{kaneBandStructureIndium1957, luttingerMotionElectronsHoles1955}
Larger band models have also been 
constructed for group-IV,
zincblende, and wurtzite semiconductors.\cite{cardonaEnergyBandStructureGermanium1966, fariajuniorRealisticMultibandApproach2016, gawareckiInvariantExpansion30band2022}
For other material classes or other high-symmetry points in the Brillouin zone, readily parametrized
\kdotp models are seldom available.

\kdotp models have a valuable role for analyzing and understanding electronic band structures,
and they can also be used as input for computing electronic properties of materials, such as dielectric response or conductivity.
Moreover, they have a foundational role in the construction of 
effective multiband Schr\"odinger equations in the envelope-function approximation,
which can be obtained through the 
substitution $\mathbf{k} \rightarrow -\mathrm{i}\nabla$, although care must be taken in operator ordering~\cite{bastardSuperlatticeBandStructure1981,burtJustificationApplyingEffectivemass1992}.
As such, the \kdotp method can bridge solid-state electronic structure calculations to the simulation of quantum states in nanostructures, such as quantum wells, wires, dots, superlattices, and more general semiconductor heterostructures.  

While the \kdotp method is mostly associated with empirical model Hamiltonians, Eq.~\eqref{eq:bare-kp} can, in principle, be evaluated directly from first principles, using $\varepsilon_n$ and
$\bm{\pi}_{nm}$ output of Kohn--Sham\cite{kohnSelfConsistentEquationsIncluding1965}   
calculations.\cite{perssonFullbandMethodSolving2007, berlandEnablingAccurateFirstprinciple2017, jocicInitioConstructionSymmetryadapted2020, v.cassianoDFT2kpEffectiveKp2024}
However, such direct
first-principles models converge slowly with the number of
bands.\cite{perssonFullbandMethodSolving2007, berlandEnablingAccurateFirstprinciple2017, v.cassianoDFT2kpEffectiveKp2024}
Truncated numerical models nevertheless remain useful for small
$\mathbf{q}$, such as for an ingredient in Brillouin-zone
interpolation schemes.\cite{perssonFullbandMethodSolving2007, berlandEnablingAccurateFirstprinciple2017, berlandThermoelectricTransportGaAs2018}
In projector-augmented wave or other pseudopotential calculations,
the potential is nonlocal,\cite{levineLinearOpticalResponse1989, pickardSecondorderPerturbationTheory2000}
and a first-principles \kdotp method becomes inexact.
The potential is also nonlocal in exact exchange used in hybrid functionals and in other methods beyond standard Kohn--Sham density functional theory (DFT).

The slow convergence with the number of bands and inexactness of raw 
 \kdotp  parameters with nonlocal potentials motivates
renormalizing the $\bm{\pi}_{nm}$ matrix to improve agreement with the DFT band structure. 
However, parameter scaling is nontrivial since eigenstates computed with DFT generally carry an arbitrary gauge; degenerate multiplets mix and singlets acquire a complex phase. 
Gauge fixing,
i.e., rotating
the numerical basis into a symmetry-adapted representation using
group theory is a viable approach,\cite{jocicInitioConstructionSymmetryadapted2020, v.cassianoDFT2kpEffectiveKp2024,zhangVASP2KPKpModels2023}
that can provide system-specific models with 
adjustable parameters.

In this study, we instead exploit the
fact that the matrices $\bm{\pi}_{nm} = (\Pi^x_{nm}, \Pi^y_{nm}, \Pi^z_{nm})$ are Hermitian, and hence their eigenvalue spectra are gauge-invariant. 
Moreover, along symmetry-related directions, the eigenvalues are
numerically identical, and renormalizing these shared eigenvalues
provides symmetry-respecting degrees of freedom that can be
optimized to match the underlying DFT band structure.
We refer to this process as covariant renormalization: it gives the same model
whichever gauge the degenerate multiplets are in, and preserves the crystal symmetry,
without the need to enforce any specific basis.

\section{Theory and implementation}

\subsection{Covariant renormalization}
\label{sec:covariant}

Models built directly from Eq.~(\ref{eq:bare-kp}) using the
generalized momentum matrix element, $\bm{\pi}_{nm}$,  computed from first principles
are here termed \emph{bare} \kdotp models.
\emph{Renormalized} models retain the form of Eq.~(\ref{eq:bare-kp}),
but adjust $\bm{\pi}_{nm}$ to compensate for the
finite number of bands and the nonlocality of the
underlying Hamiltonian. 

The momentum matrix elements $\bm{\pi}_{nm}$ form
three 
 Hermitian $N \times N$ matrices
$\Pi^\alpha$, one for each Cartesian component 
$\alpha \in \{x,y,z\}$, where $N$ is the number of bands retained.
The gauge of these matrices can be changed under 
\begin{equation}
\Pi^\alpha \rightarrow U^\dagger \Pi^\alpha U,
\label{eq:gauge}
\end{equation}
where $U$ is a unitary block-diagonal matrix over the degenerate
multiplets.

The individual $\Pi^\alpha$ matrices allow a spectral decomposition
\begin{equation}
\Pi^\alpha = Q^\alpha \Lambda^\alpha (Q^\alpha)^\dagger,
\qquad
\Lambda^\alpha = \mathrm{diag}(\lambda^\alpha_1, \lambda^\alpha_2, \ldots),
\label{eq:spectral}
\end{equation}
with real (gauge-invariant) eigenvalues $\lambda^\alpha_i$.
Critically, individual $\Pi^\alpha$ matrices related by symmetry
must share the same eigenvalue spectrum (up to a sign).
These recurring magnitudes 
identified within a numerical tolerance define magnitude sets
that can be scaled with the same dimensionless parameter $\eta_g$,
\begin{equation}
\lambda^\alpha_i \rightarrow
\lambda^\alpha_i \left[1 + \eta_{g(|\lambda^\alpha_i|)}\right],
\label{eq:renorm}
\end{equation}
with $\tilde{\Pi}^\alpha$ reconstructed from Eq.~\eqref{eq:spectral}, hence $\eta_g = 0$ recovers the bare
model. 
 Near-null eigenvalues are
discarded to limit numerical noise.

When the equivalent directions are mutually perpendicular, so that they can be taken
as Cartesian axes, like at the $\Gamma$ point of cubic, tetragonal, and orthorhombic crystals,
Eq.~\eqref{eq:renorm} is applied to each $\Pi^\alpha$ directly, with the same
$\eta_g$ for equivalent axes.
When the momentum matrices along symmetry-related directions are instead
linearly dependent, rescaling them with the same $\eta_g$ gives matrices with the
rescaled spectra that in general no longer satisfy the same linear relation,
since rescaling a sum of matrices does not give the sum of the rescaled matrices.
This is the case at the $\Gamma$ point of hexagonal and trigonal crystals, where the
directions equivalent to $x$ are $60^\circ$ and $120^\circ$ away in the plane.
With $\Pi_\theta = \cos\theta\,\Pi^x + \sin\theta\,\Pi^y$, the matrices along
$\theta = 0^\circ$, $60^\circ$, and $120^\circ$ satisfy
\begin{equation}
\Pi_{0^\circ} - \Pi_{60^\circ} + \Pi_{120^\circ} = 0 ,
\label{eq:relation}
\end{equation}
since $\cos0^\circ - \cos60^\circ + \cos120^\circ = 0$, and likewise for the sines.
The rescaled matrices $F_\theta$ in general do not satisfy Eq.~\eqref{eq:relation}.
We therefore project them onto matrices that do
(detailed in Appendix~\ref{app:projection}),
\begin{equation}
\begin{aligned}
X_{0^\circ} &= F_{0^\circ} - \tfrac13 S, \\
X_{60^\circ} &= F_{60^\circ} + \tfrac13 S, \\
X_{120^\circ} &= F_{120^\circ} - \tfrac13 S ,
\end{aligned}
\label{eq:projection}
\end{equation}
with $S = F_{0^\circ} - F_{60^\circ} + F_{120^\circ}$.
The renormalized Cartesian components follow from
$\Pi^x = \Pi_{0^\circ}$ and $\Pi^y = (\Pi_{60^\circ} + \Pi_{120^\circ})/\sqrt3$,
with $X_\theta$ in place of $\Pi_\theta$,
\begin{equation}
\begin{aligned}
\tilde\Pi^x &= X_{0^\circ}
 = \tfrac23\left(F_{0^\circ} + \tfrac12 F_{60^\circ} - \tfrac12 F_{120^\circ}\right), \\
\tilde\Pi^y &= \tfrac{1}{\sqrt3}\left(X_{60^\circ} + X_{120^\circ}\right)
 = \tfrac{1}{\sqrt3}\left(F_{60^\circ} + F_{120^\circ}\right),
\end{aligned}
\label{eq:hexagonal}
\end{equation}
which reduce to $\Pi^x$ and $\Pi^y$ for $\eta_g = 0$.
The $c$-axis component $\Pi^z$ is renormalized separately.

The number of free parameters in this covariant renormalization is far lower
than the number of matrix elements.
When $\Pi^x$, $\Pi^y$, and $\Pi^z$ are related by symmetry, as at the $\Gamma$ point
of a cubic crystal, it equals the number of distinct eigenvalue magnitudes of $\Pi^x$.
At the $\Gamma$ point of a hexagonal crystal, it is the sum of the distinct
magnitudes of $\Pi_{0^\circ}$ and of $\Pi^z$.

\subsection{Downfolding}
\label{sec:downfolding}

For basic analysis or as input to envelope function models,
compact few-band models with good performance close to a given high-symmetry point can be both sufficient and desirable.
Such models can be constructed from a larger model with downfolding.

In downfolding, remote bands are accounted for perturbatively through L\"owdin
partitioning \cite{lowdinNoteQuantumMechanicalPerturbation1951, winklerQuasiDegeneratePerturbationTheory2003}.
The Hamiltonian containing 
 $N$ bands is partitioned in terms of a center block $\mathcal{C}$ of $n_c$ bands 
 and a remote part $\mathcal{R}$, in a block form,
 \begin{equation}
H(\mathbf{k}) =
\begin{pmatrix}
H_{\mathcal{C}}(\mathbf{k}) & h(\mathbf{k}) \\
h^\dagger(\mathbf{k}) & H_{\mathcal{R}}(\mathbf{k})
\end{pmatrix}\,,
\label{eq:block}
\end{equation}
The off-diagonal block takes the form 
$h(\mathbf{k}) = (\hbar/m) \sum_\alpha q_\alpha
\Pi^\alpha_{\mathcal{C}\mathcal{R}}$, where
$\Pi^\alpha_{\mathcal{C}\mathcal{R}}$ is the off-diagonal block of
$\Pi^\alpha$.
In the partitioning, it is critical to not split degenerate multiplets,
requiring care in numerical implementation.
Eliminating the remote
block from the eigenvalue problem gives
\begin{equation}
H^{\mathrm{eff}}(E, \mathbf{k}) = H_{\mathcal{C}}(\mathbf{k})
+ h(\mathbf{k}) \left[ E - H_{\mathcal{R}}(\mathbf{k}) \right]^{-1}
h^\dagger(\mathbf{k}),
\label{eq:schur}
\end{equation}
which reproduces the eigenvalues of the full Hamiltonian exactly, but depends on
the eigenvalue $E$ and hence must be solved self-consistently.
Successive orders of downfolding follow from expanding the
resolvent.
To second order in the coupling, with the resolvent
evaluated at the energies at $\mathbf k_0$ and symmetrized to preserve hermiticity, the downfolded model takes
the closed form
\begin{equation}
H^{\mathrm{eff}}(\mathbf{k}) = E_{\mathcal{C}}
+ \frac{\hbar}{m} \sum_\alpha q_\alpha \Pi^\alpha_{\mathcal{C}\mathcal{C}}
+ \sum_{\alpha\beta} q_\alpha q_\beta\, D^{\alpha\beta},
\label{eq:quad}
\end{equation}
with $E_{\mathcal{C}}$ the diagonal matrix of center energies,
$\Pi^\alpha_{\mathcal{C}\mathcal{C}}$ the center block of
$\Pi^\alpha$, and the curvature tensor
\begin{align}
D^{\alpha\beta} ={}& \frac{\hbar^2}{2m}\, \delta^{\alpha\beta}\, I
\nonumber \\
&+ \frac{\hbar^2}{4m^2}
\left( K^\alpha (\Pi^\beta_{\mathcal{C}\mathcal{R}})^\dagger
+ K^\beta (\Pi^\alpha_{\mathcal{C}\mathcal{R}})^\dagger
+ \mathrm{H.c.} \right),
\label{eq:dtensor}
\end{align}
where
$K^\alpha_{nm} = \pi^\alpha_{nm}/(\varepsilon_n - \varepsilon_m)$,
with $n \in \mathcal{C}$ and $m \in \mathcal{R}$, is the
first-order mixing of remote into center states. The free-electron
contribution is here carried inside $D^{\alpha\beta}$, so no separate
kinetic term appears in Eq.~\eqref{eq:quad}.

If only a single band is retained as the center block, we obtain simply an effective mass,
such as that of a conduction-band electron in III-V semiconductors. 
Retaining instead a multiplet, such as the valence multiplet, Eq.~\eqref{eq:dtensor} 
gives a curvature matrix in the Luttinger and Kohn form.\cite{luttingerMotionElectronsHoles1955}
Keeping also the conduction band inside the center block instead gives a 4-band Kane 
model or 8-band with spin-orbit coupling used traditionally for narrow-gap
semiconductors.\cite{kaneBandStructureIndium1957}

Renormalization of \kdotp models can also be done either using a larger-band model, 
or directly on a downfolded form. 
An option is to covariantly renormalize
$D^{\alpha\beta}$ directly,
which, after a certain number of bands in the remote block, will result in fewer adjustable parameters than adjusting 
$\Pi^\alpha$ of the full Hamiltonian. 
However, here we choose to simply renormalize
the spectrum of the full $\Pi^\alpha$, 
allowing us to treat the eigenvalues that enter the $D^{\alpha\beta}$ and those in the center block 
on the same footing.  
When many remote bands are used, there will be a redundancy in the determination of $D^{\alpha\beta}$, but this  
only changes numerical optimization details and should have limited bearing on the final models. 


\subsection{Computational implementation}

DFT calculations were performed with the planewave projector augmented-wave (PAW) method in the 
VASP software package \cite{blochlProjectorAugmentedwaveMethod1994, kresseEfficiencyAbinitioTotal1996, vasp4, kresse1999}.
The matrix elements $\bm{\pi}_{nm}$ and band velocities were computed from
wave function derivatives with respect to $\mathbf{k}$, in the same manner as
for standard independent-particle dielectric
properties.\cite{readCalculationOpticalMatrix1991,gajdosLinearOpticalProperties2006a, berlandThermoelectricTransportGaAs2018}
This evaluates the velocity operator of Eq.~\eqref{eq:pi} directly, and since
the derivatives are taken within the self-consistent calculation itself,
spin-orbit coupling enters $\hat{\bm{\pi}}$ whenever it is included in the DFT calculation.

Lattice constants were optimized with the vdW-DF-cx functional,\cite{behy14}
chosen due to
its generally accurate lattice constants for solids \cite{tranNonlocalVanWaals2019} while also providing a physically unified treatment should nonlocal correlation become relevant.
Electronic properties and band structures were computed using the PBE generalized gradient approximation functional~\cite{pbe1996}.
This functional generally underestimates band gaps compared to experiment,
and \kdotp models for practical use might require more accurate methods such as hybrid functionals or GW calculations. 
The planewave cutoff was set to 1.3 times the largest recommended cutoff energy among the constituent PAW datasets, rounded up to the nearest 10 eV, for both structural optimization and electronic-structure calculations.
For relaxation and charge density, a k-spacing of 0.2 \AA$^{-1}$ was chosen. 
As reference data for the \kdotp models designed for particular high-symmetry points,
we computed Kohn--Sham eigenvalues on rays out of the
high-symmetry point. The rays are taken along the $\langle 100\rangle$,
$\langle 110\rangle$, and $\langle 111\rangle$ direction families of the
primitive reciprocal basis, reduced to symmetry-inequivalent
representatives by the crystal symmetry and time reversal and sampled on 21 equidistant points per ray.
Each ray extends a fraction $f$ of the primitive reciprocal lattice vectors along its
direction indices, allowing the same prescription to be applied for any crystal symmetry or lattice constant. 
In this fashion the span of the rays in the different directions of the first Brillouin zone can vary significantly;
for GaP at $f = 0.2$ the rays reach $40\%$ of the way to L and to X, and $53\%$ of the way to K.
The extent is chosen per material: $f = 0.2$ for GaP, $0.3$ for zincblende AlN and for PbTe,
and $0.4$ for wurtzite AlN, the larger values providing more weight to nonparabolic features.
However, the data used in the fitting is set mostly by an energy window, 
set to 0.3~eV below the valence- and 0.3~eV above the conduction-band energies at the
high-symmetry point in most cases, but expanded to 0.4~eV for wurtzite AlN.  

The loss function for determining the renormalization parameters is given by 
\begin{equation}
L(\bm{\eta}) = (1 - \omega) L_E(\bm{\eta})
+ \omega L_V(\bm{\eta})
+ \frac{\mu}{N_\eta} \|\bm{\eta}\|^2,
\label{eq:loss}
\end{equation}
where $L_E$ and $L_V$ are root-mean-square deviations from the reference
eigenvalues and band velocities.
Energy deviations are normalized by the largest energy change of the fitted bands along the rays,
and velocity deviations are taken relative to $|v| + 0.1$~\AA$^{-1}$, which weights the small
velocities near $\mathbf k_0$ that set the effective masses.
Velocities are summed over degenerate multiplets, where only the sum is gauge invariant.
The ridge term constrains the model from deviating strongly
from the bare one. The division by $N_\eta$ is made to allow comparable regularization terms, $\mu$, 
across basis sizes, 
as the sum scales with $N_\eta$.
The loss function is minimized with the
limited-memory BFGS method~\cite{liuLimitedMemoryBFGS1989}, as implemented in the
\textsc{Optim.jl} package~\cite{kmogensenOptimMathematicalOptimization2018}, with gradients evaluated
by central finite differences.
The loss \eqref{eq:loss} is evaluated with 
$\omega = 0.5$, and the regularization parameter was set to  
$\mu = 4\times10^{-3}$ throughout. 


The density of states was computed with the tetrahedron method. The electronic conductivity was  computed directly without any interpolation,\cite{scheidemantelTransportCoefficientsFirstprinciples2003}
with velocities obtained directly from DFT or with \kdotp through a unitary evolution of the $\bm{\pi}_{nm}$
matrix to the given $\mathbf{k}$-point. 
This scheme is unlike, e.g., BoltzTraP~\cite{madsenBoltzTraPCodeCalculating2006,madsenBoltzTraP2ProgramInterpolating2018},
which uses Fourier interpolation. 
The \kdotp renormalization, downfolding, and Brillouin-zone integrations
were performed with \textsc{StateCraft}, a Julia framework for
electronic-structure analysis developed by the author.
\begin{figure}[t!]
\setlength{\abovecaptionskip}{2pt}
\setlength{\belowcaptionskip}{6pt}
\includegraphics[width=\columnwidth]{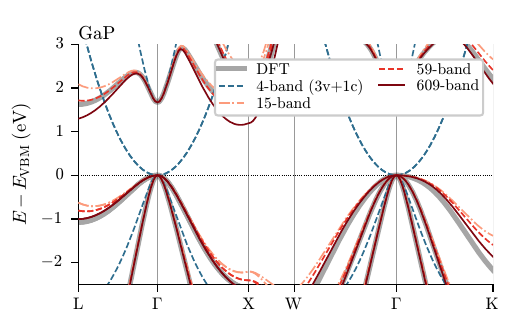}
\caption{Bare \kdotp band structures of GaP plotted against the DFT reference bands.
	The 4-band model includes the valence-band triplet
	and one conduction band; the others start from the lowest state. 
\label{fig:gap-bands-vs-basis}}
\includegraphics[width=\columnwidth]{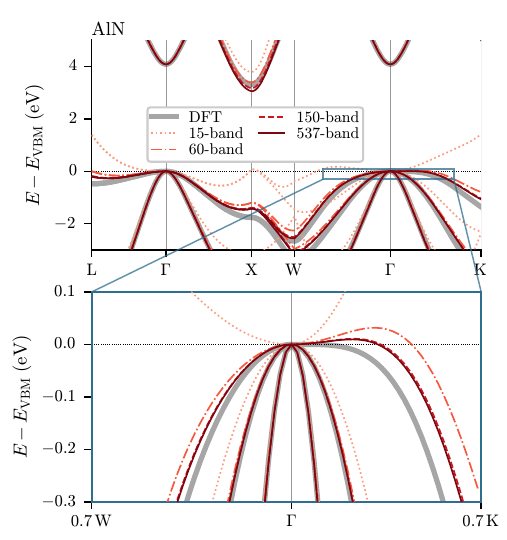}
\caption{ Bare \kdotp  vs DFT for zincblende AlN band structures, with a zoom (lower panel), extending $70\%$ of the way out to W and to K about the $\Gamma$ valence-band maximum. 
\label{fig:warping-wgk}}
\end{figure}

\section{Results}


\subsection{Issues with bare \kdotp models}
Figure~\ref{fig:gap-bands-vs-basis} shows bare \kdotp bands of GaP obtained with four 
different basis sizes as given by the number of bands included in the model, compared with the DFT reference. 
The results of the \kdotp model improve rapidly with number of bands at first, but the agreement also worsens in some respects
as the number of bands approaches the largest meaningful maximum of
609 bands for the chosen energy cutoff. The lack of appreciable improvement beyond a certain number of bands
can be due to the nonlocality of the PAW potentials, but also due to numerical inaccuracies. 


Figure~\ref{fig:warping-wgk} shows a corresponding comparison for zincblende AlN.
Here, overall features do improve with basis size, at least up to around 150 bands;
however, no matter how many bands are included, the bare \kdotp method
fails to provide the correct band curvature, i.e., the sign of the effective mass, in the $[110]$ direction.
The near-gap valence-band structure of this material is highly warped, with a very high valence-band effective mass in the  $[110]$ direction, unlike in the other directions.


\subsection{Renormalized many-band \kdotp models}
\begin{figure}[t!]
\includegraphics[width=\columnwidth]{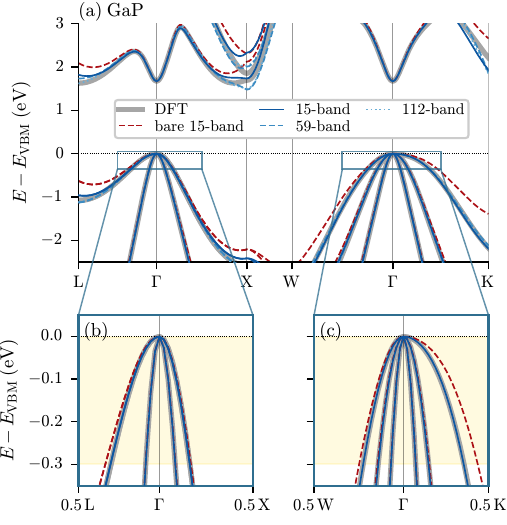}
\caption{Renormalized \kdotp bands of GaP compared with the bare 15-band model. 
The full path (a) and magnified (b), (c) close to the VBM at the two $\Gamma$ passages.
The shade indicates the 0.3-eV energy window of the fit.
Only 15-band models are displayed in (b) and (c).
\label{fig:gap-renorm-bands}}
\end{figure}


Renormalized many-band \kdotp models massively improve agreement with DFT compared to the bare.
We first demonstrate this finding for models targeting high performance around specific high-symmetry points.

\subsubsection{GaP}

Figure~\ref{fig:gap-renorm-bands} compares \kdotp renormalized models
with the bare 15-band model and the DFT reference. The renormalized models appear virtually indistinguishable
from the DFT within the energy range used for fitting.


\begin{figure}[h!]
\includegraphics[width=\columnwidth]{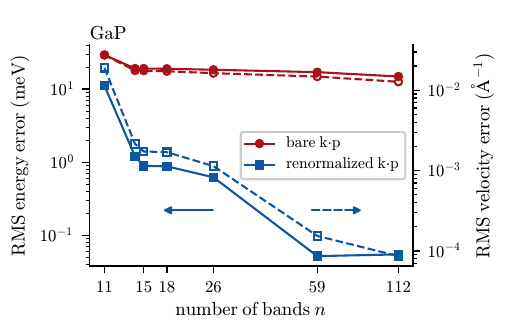}
\caption{RMS errors of \kdotp models within the fitting regime. The
left vertical axis indicates energy errors, while the right indicates velocity errors (see arrows). 
\label{fig:gap-renorm-rmse}}
\end{figure}

Figure~\ref{fig:gap-renorm-rmse} and Table~\ref{tab:gap-renorm-rmse} 
provide the corresponding errors and values of the loss function, Eq.~\eqref{eq:loss}. 
They reveal that the renormalized models, unlike the bare models, continue to improve with the number of bands in the basis. 
The precise number of bands used is selected in part to ensure a certain energy distance between the degenerate multiplets. 


\begin{table}
\caption{Fitting accuracy vs number of bands. $\Delta E$ (meV)
and $\Delta v$ ($10^{-3}$\,\AA$^{-1}$) are the energy and group-velocity errors,
the latter expressed as $m v/\hbar$, and
$N_\eta$ the number of renormalization parameters. $L$ ($10^{-3}$) is the
loss of the renormalized model, Eq.~\eqref{eq:loss}, provided for the data term alone
and with the ridge term included.
\label{tab:gap-renorm-rmse}}
\begin{tabular}{rr@{\hskip 1.4em}rr@{\hskip 1.4em}rr@{\hskip 1.4em}rr}
\toprule
 & & \multicolumn{2}{c}{$\Delta E$} & \multicolumn{2}{c}{$\Delta v$} & \multicolumn{2}{c}{$L$} \\
\cmidrule(lr){3-4}\cmidrule(lr){5-6}\cmidrule(lr){7-8}
$n$ & $N_\eta$ & bare & ren. & bare & ren. & data & full \\
\midrule
11 & 3 & 29.4 & 11.16 & 27.5 & 19.00 & 26.53 & 27.67 \\
14 & 4 & 19.2 & 1.21 & 17.6 & 2.15 & 2.55 & 2.56 \\
15 & 4 & 19.1 & 0.89 & 17.5 & 1.73 & 3.22 & 3.23 \\
18 & 6 & 19.0 & 0.88 & 17.3 & 1.69 & 3.26 & 3.27 \\
26 & 9 & 18.4 & 0.62 & 16.3 & 1.14 & 2.60 & 2.71 \\
59 & 21 & 17.1 & 0.05 & 14.8 & 0.15 & 0.18 & 0.21 \\
112 & 40 & 14.9 & 0.06 & 12.7 & 0.09 & 0.12 & 0.16 \\
\bottomrule
\end{tabular}

\end{table}

\begin{figure}[t!]
\includegraphics[width=\columnwidth]{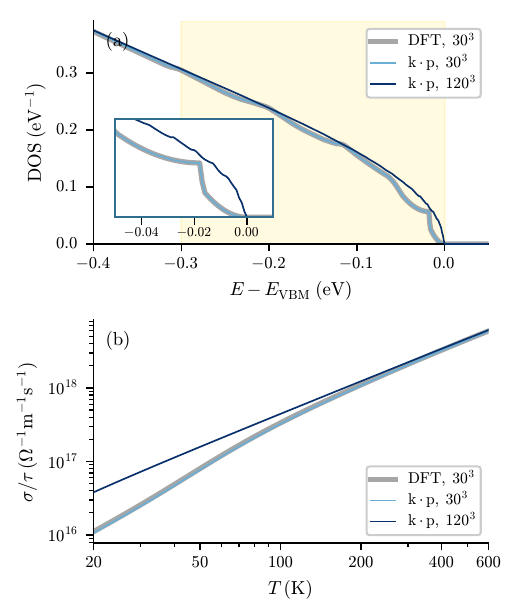}
\caption{Renormalized 59-band GaP model against
the DFT reference and on a dense mesh. (a) Valence density of states, magnified at the onset;
(b) hole conductivity with the Fermi level at the valence-band maximum,
at a constant relaxation time. 
\label{fig:gap-bz-integrals}}
\end{figure}

Once an accurate model is constructed, it can be evaluated on meshes far
denser than the reference calculation it was fitted to, as shown in Fig.~\ref{fig:gap-bz-integrals}.
In panel a, it shows how the \kdotp fit perfectly reproduces the DOS computed with a separate DFT calculation, including the artifacts of a tetrahedron integration. Using the same model, we also perform a \kdotp calculation on a $120^3$ mesh,
this calculation also exhibiting minute tetrahedron integration wiggles evidenced in the inset. 
In panel b, we show the conductivity computed with the Fermi level set to the VBM edge as a function of temperature. 
It shows how the \kdotp model follows the DFT curve closely when evaluated on the same mesh, but there is a significant disparity at lower temperature with the same \kdotp model evaluated on a denser mesh. 
The result underlines the high k-sampling that can be required for transport calculations at low temperature and that the \kdotp method can be used for this purpose. 

\subsubsection{Zincblende AlN}

\begin{figure}[t!]
\includegraphics[width=\columnwidth]{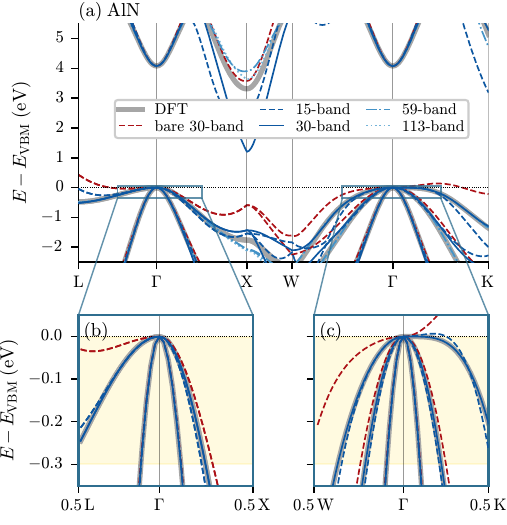}
\caption{Renormalized \kdotp bands of zincblende AlN optimized in the near-gap $\Gamma$-region 
(a). The two zoom-in regions (b), (c), show respectively two $\Gamma$ passages in the valence-band region; the shade indicates the energy window of the fit.
In the zoom-in region, only the bare 30-band model and
the renormalized 15 and 30-band models are drawn, the larger renormalized models
being indistinguishable from the 30-band model at this scale.
\label{fig:aln-renorm-bands}}
\end{figure}

Fig.~\ref{fig:aln-renorm-bands} compares the bare 30-band model for AlN
with different renormalized models.  
While improving agreement with DFT, the renormalized 15-band model
fails to provide the correct band curvature in the [110] direction; however, extending the basis to 30 bands 
does resolve the issue.

\subsubsection{Wurtzite AlN}

\begin{figure}[t!]
\includegraphics[width=\columnwidth]{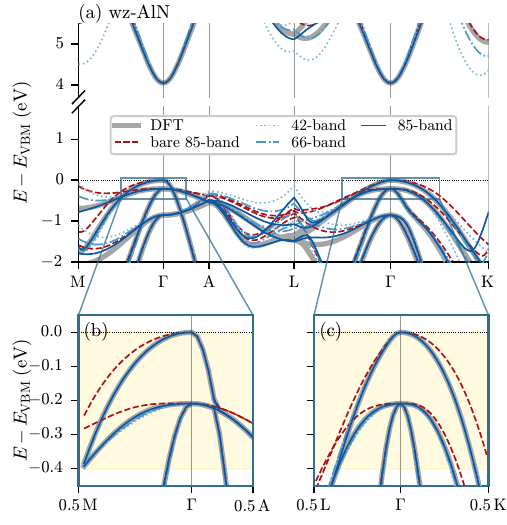}
\caption{Renormalized \kdotp bands of wurtzite AlN along the
M--$\Gamma$--A--L--$\Gamma$--K path (a), with the bare 85-band model shown for
reference. Panels (b) and (c) magnify the valence-band top at the two $\Gamma$
passages, along M--$\Gamma$--A and L--$\Gamma$--K respectively; the shade indicates
the 0.4-eV energy window of the fit. The zooms show the 42-, 66- and 85-band models;
excellent agreement requires an 85-band model.
\label{fig:alnwz-renorm-bands}}
\end{figure}

To test the  \kdotp renormalization scheme for a system with lower symmetry, 
we consider wurtzite AlN. 
Since the DFT reference data exhibit a crystal-field splitting creating a doublet 0.21~eV below the VBM, the fitting energy span was expanded to 0.4 eV below the VBM,
so that all three valence bands fall comfortably within the fit window.
Similar to the case of zincblende AlN, a considerable number of bands are needed
to obtain an accurate model, as shown in Fig.~\ref{fig:alnwz-renorm-bands}.
The zoomed-in region shows that a renormalized 
85-band model provides excellent agreement with DFT near the $\Gamma$-point, 
while the 42- and 66-band models remain visibly off (the latter only with high-resolution digital reading). 
A 106-band model brings no appreciable further improvement.
The 85-band model relied on 56 renormalization parameters, 29 for the in-plane and 27
for the $c$-axis components.
Its band energies along directions related by the sixfold rotation agree to
within 0.1~meV on the reference rays, as for the bare model.

\subsubsection{PbTe: \texorpdfstring{$L$}{L}-centered model with spin-orbit coupling}

\begin{figure}[t!]
\includegraphics[width=\columnwidth]{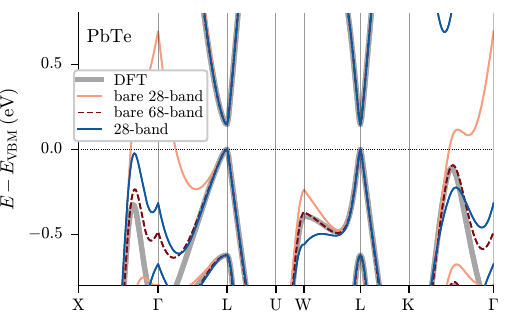}
\caption{
	Band structures of rocksalt PbTe for bare 28- and 68-band \kdotp models and renormalized 28-band \kdotp model (blue) plotted against the DFT reference, expanded at the $L$ point with spin-orbit coupling included. 
\label{fig:pbte-showcase}}
\end{figure}

The renormalization scheme can also be readily extended to 
other high-symmetry points, as shown for PbTe in Fig.~\ref{fig:pbte-showcase}.
It compares a bare 28-band and 68-band model with a renormalized 28-band model.
For this system, a bare 68-band model does exhibit good agreement with DFT in the near-gap region, unlike a 28-band model. Near the $L$ point, however, a renormalized 28-band model fitted to this region does equally well.

PbTe includes strong spin-orbit effects and the comparison shows that the renormalization scheme generalizes to such systems, without the need for measures beyond turning on spin-orbit in the underlying VASP calculation. 
Spin-orbit coupling doubles the number of bands required, hence a 28-band model can be considered relatively compact.

\section{Compact few-band models}
\label{sec:compact}

\begin{figure}[t!]
\includegraphics[width=\columnwidth]{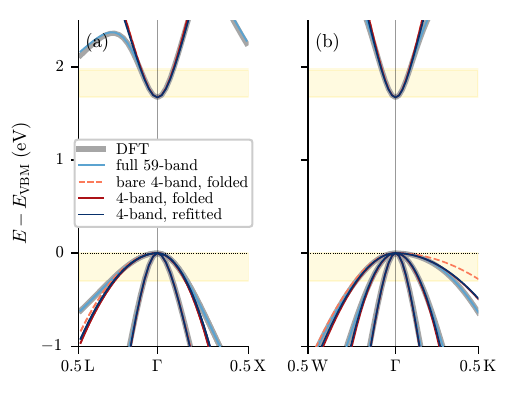}
\caption{Four-band models of GaP obtained by downfolding, against the
renormalized 59-band model they are folded from, at the two $\Gamma$ passages of
the path: L--$\Gamma$--X in (a) and W--$\Gamma$--K in (b), the latter carrying
the heavy $[110]$ valence direction. The bare four-band model is folded from the
first-principles energies and momentum matrix elements directly. 
The two others are renormalized, one reoptimized after folding. The
shade indicates the 0.3-eV energy window of the fit.
\label{fig:gap-downfold}}
\end{figure}

So far we have seen how 
renormalized models can obtain excellent agreement with DFT band structure, but sufficient agreement may also require a significant number of bands. 
While a 15 or even 60-ish band model may be of significant practical utility for instance for computing transport properties,
as input to envelope functions or as tabulated data for comparison with experiment, more compact models are desirable. 
Such models can be obtained with downfolding as described in Sec.~\ref{sec:downfolding}.
Here we illustrate this downfolding procedure for GaP. We start from a very accurate renormalized 59-band model and downfold it to a 4-band model of 3 valence and 1 conduction bands.
Since downfolding introduces a perturbative approximation, we also make a model where we reoptimize $\bm{\eta}$
in its downfolded form. Figure~\ref{fig:gap-downfold} shows both, together with the
four-band model obtained by downfolding with the bare parameters.

First, the downfolded bare model follows the reference closely along several directions but fails
along $[110]$, where the valence bands are heaviest. 
Folding the renormalized model removes the issue.  
Reoptimizing the downfolded model improves the agreement further, though by less.
The reoptimization uses the same rays and 0.3-eV energy window as the 59-band model,
restricted to $|\mathbf q| \le 0.15$~\AA$^{-1}$, about 15\% of the way to L and 12\% to K,
since a four-band model cannot follow the bands much further out.
Within this region, the root-mean-square energy deviation falls from 4.2 to 3.9 to
2.4~meV across the three models. Over the full rays, which reach 40\% of the way to
L and X and 53\% to K, it falls from 26.9 to 15.1 to 11.5~meV.
The reoptimized model reaches this at $\max|\eta| = 0.34$, slightly below the
$0.37$ of the 59-band model it starts from, so the compact model is not bought
by straying further from the first-principles matrix elements.

\section{Full-zone models}
\label{sec:fullzone}

\begin{figure*}[t!]
\includegraphics[width=\textwidth]{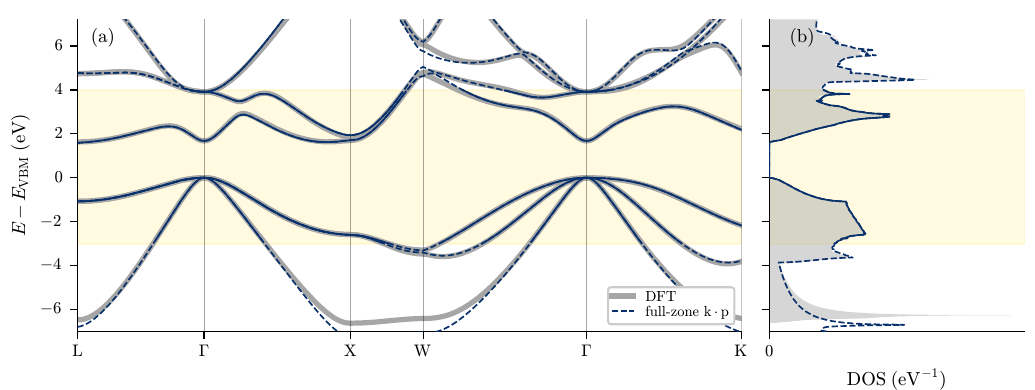}
\caption{Band structure (a) and density of states (b) of GaP from a 59-band
model renormalized over the whole Brillouin zone, against the DFT reference.
The gold band marks the energy window the model was fitted in, and the model is
drawn dashed outside it. The fit used the irreducible wedge of a regular
$30^3$ mesh, so the band path is held out. The reference density of states is
shaded; both are integrated on the same mesh.
\label{fig:gap-fullzone}}
\end{figure*}

The models considered so far are renormalized against rays out of a
high-symmetry point, and are hence designed to accurately describe the region close to this point. 
Here we replace the rays with the points of the irreducible Brillouin zone of a
$30^3$ mesh. Each point is evaluated at the shortest $\bm{q}$, shifting it by a
reciprocal-lattice vector $\bm{G}$ where this brings it closer to $\mathbf k_0$. The fitting energy window is set from
$-3$ to $4$~eV about the valence-band maximum.

Figure~\ref{fig:gap-fullzone} compares DFT and the \kdotp model for the 
band structure and DOS. Visually, it exhibits a remarkable agreement between the two within the fitting regime, and overall good agreement beyond, except for the band flattening 4~eV below the lower edge of the fitting window.   
However, while appearing accurate at this scale, full-zone models are less accurate 
than the models targeting narrower regimes.

\section{Discussion}

This paper establishes a covariant renormalization of \kdotp Hamiltonians to reproduce electronic band structures. 
We have demonstrated how accurate \kdotp models can be constructed across fairly diverse semiconductor materials,
and we have seen how AlN systems require models with a far larger number of bands than GaP. 
The inclusion of AlN as a test system was made precisely because it emphasized this challenge.
We speculate that first-principles \kdotp models are simpler to construct for low-bandgap systems with localized orbitals, 
e.g., d orbitals, located far from the band edges.

We also showcased how different types of \kdotp model can be constructed from first principles. 
Our default model was a many-band model optimized around a high-symmetry point. While we generally emphasized the smallest accurate model
in the visualization, e.g., a 15-band model for GaP, in practical applications one may opt for a larger model since computational costs are modest,
say for evaluating a 59-band model, 
as was done for 
 generating a finely sampled DOS and for the transport calculations.
For compact few-band models, it is critical to incorporate remote bands, which can be done through downfolding. 
Another class entirely is the full-zone models, which are fitted across the entire Brillouin zone; 
we showed how this was feasible for GaP. 

The work provides a natural starting point for constructing databases of \kdotp models that can be adapted for different purposes, 
e.g., as input to semi-classical transport modelling, or modelling of nanostructures with envelope functions, or simply as compact models that can be used for comparison with experimental spectroscopy.  
However, the number of bands required and the natural fitting cutoffs (e.g., full zone, effective masses) will differ with material system and intended usage. 
In such an effort, higher-level methods such as hybrid functionals and GW would be natural. In a practical workflow, an option is to compute momentum matrix elements at a lower level of theory and renormalize them to the higher level of theory, easing the regularization requirements.

\appendix
\section{Projection onto the in-plane relation}
\label{app:projection}

We collect the three rescaled in-plane matrices of Sec.~\ref{sec:covariant} as a
column over directions, $\mathbf F = (F_{0^\circ}, F_{60^\circ}, F_{120^\circ})^{\mathsf T}$,
and write the coefficients of Eq.~\eqref{eq:relation} as the unit vector
$c = (1, -1, 1)^{\mathsf T}/\sqrt3$, with $c^{\mathsf T}c = 1$.
Equation~\eqref{eq:relation} then reads $c^{\mathsf T}\mathbf X = 0$, while
$c^{\mathsf T}\mathbf F = S/\sqrt3$.
Removing the component of $\mathbf F$ along $c$,
\begin{equation}
\mathbf X = P\,\mathbf F, \qquad
P = 1 - c\,c^{\mathsf T}
= \frac13
\begin{pmatrix}
 2 & 1 & -1 \\
 1 & 2 &  1 \\
-1 & 1 &  2
\end{pmatrix} ,
\end{equation}
gives Eq.~\eqref{eq:projection} row by row, as
$X_\theta = F_\theta - c_\theta\, S/\sqrt3$ with $c_\theta = \pm 1/\sqrt3$.
$P$ acts on the direction index only.
The relation holds,
$c^{\mathsf T}\mathbf X = c^{\mathsf T}\mathbf F - (c^{\mathsf T}c)(c^{\mathsf T}\mathbf F) = 0$.

\begin{acknowledgments}
This work was funded in part by the Research Council of Norway through the MORTY (315330)
and NOMATEC (360069) projects.
\end{acknowledgments}

\section*{Data availability}

A snapshot of the \textsc{StateCraft} code, together with the scripts, DFT band
data, and fitted models used in this work, is deposited at Zenodo, with access
available upon reasonable request.\cite{statecraft_snapshot}

\bibliography{references}

\end{document}